# SPIN FLUCTUATIONS AND "SPIN NOISE"

SANDIPAN PRAMANIK

*Department of Electrical Engineering, Virginia Commonwealth University, Richmond, VA 23284, USA*

SUPRIYO BANDYOPADHYAY

*Department of Electrical Engineering and Department of Physics, Virginia Commonwealth University, Richmond, VA 23284, USA*

We have theoretically studied the temporal fluctuations and the resulting kinetic noise in the average spin polarization of an electron ensemble drifting in a quantum wire under a high electric field. Electrons are initially injected in the wire from a ferromagnetic contact with all their spins polarized along the wire axis. The average spin polarization of the ensemble decays during transport because of D'yakonov-Perel' relaxation caused by both Rashba and Dresselhaus interactions. Once steady state is reached, the average spin fluctuates randomly around zero. The time average of this fluctuation is zero. The autocorrelation function of this fluctuation approximates a Lorentzian and so does the spectral density. To our knowledge, this is the first study of spin fluctuations and "spin noise" in a nanostructure.

1. Introduction

"Spintronics" is a rapidly burgeoning field of science and technology dedicated to the development of electronic and optical devices that exploit the spin degree of freedom of charge carriers to elicit a myriad of storage and processing functions. A number of spintronic device proposals have appeared in the literature, e.g. spin-HEMTs[1], diodes[2], solar cells[3], filters[4], stub tuners[5], spin-coherent photodetectors[6] and quantum spin field effect transistors that transport a spin current in the absence of any charge current[7]. Modeling spin transport in *nanostructures* has gained particular importance since quantum confinement has been found to strongly suppress spin depolarization mechanisms, making spin polarization long-lived.

Spin transport in semiconductors has been treated in the past with a variety of models such as: (1) *single particle ballistic models* which are fully quantum mechanical but do not account for any scattering or spin dephasing effects[8,9]; (2) *phase coherent quantum mechanical approaches* that treat spin dephasing via elastic scattering only[10]. These are more sophisticated than the ballistic models but do not account for inelastic (or phase-breaking) scattering mechanisms, which are important at high temperatures and electric fields; (3) *linear classical drift diffusion models*[11,12] that cannot handle non-linear and non-local effects. More importantly, being classical, they cannot account for interference between orthogonal spin states (e.g. "spin-up" and "spin-down" states); and finally (4) *semi-classical non-linear models*[13-16] which couple spin density matrix evolution (based on a fully quantum mechanical Sturm-Liouville type equation) with the semi-classical Boltzmann transport equation. This approach can account for non-linear transport effects, as well as interference effects between orthogonal spin states. So far, these models have revealed surprising features of spin transport in semiconductor quantum wires. For instance, it has been shown that spin relaxation rate can be very anisotropic (spin injected along the wire is much longer lived than spin injected transverse to the wire axis) and the relaxation rate can be suppressed by at least an order of magnitude by quasi one-dimensional quantum confinement[13-15]. These models are ideal for studying spin relaxation due to D'yakonov-Perel'[17] and Elliott-Yafet[18] mechanisms. They are also capable of producing information about spin fluctuations since they are microscopic and deal with a *spin distribution function* unlike the drift diffusion models that deal only with ensemble averaged "moments" of the distribution function. In this paper, we have used such a model to study temporal spin fluctuations of a steady state electron ensemble drifting in a quantum wire under a high electric field when hot carrier effects (non-local and non-linear effects) are important. To our knowledge, this is the first study of spin fluctuation and noise in a nanostructure. We present results pertaining to the autocorrelation function of the fluctuations as well as the spectral density of the associated kinetic "spin noise".

## 2. Theory

We consider a "spin-valve" type quantum wire structure consisting of a GaAs quantum wire capped by half-metallic ferromagnetic contacts with 100% spin polarization. The wire has a rectangular cross section of 30 nm × 4 nm. The confining potential in the wire is slightly asymmetric which gives rise to a uniform electric field of 100 kV/cm transverse to the wire axis (y-axis, see Figure 1).

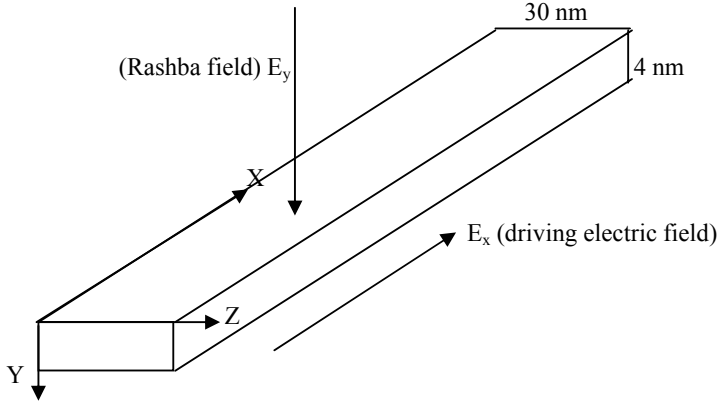

Figure 1. Geometrical structure of the quantum wire and axis designations.

This field, which can be applied by an external gate contact as visualized in reference 1, causes a Rashba interaction in the wire, but does not perturb the energy levels in the wire significantly[14]. The ferromagnetic contact is magnetized along the axis of the wire and therefore injects electrons with their spins polarized along the wire. An electric field $E_x \hat{x}$ is applied along the axis of the wire ($\hat{x}$ axis) to drive transport. As the electrons traverse the wire, they experience momentum dependent spin-orbit coupling interactions due to the Rashba effect (structural inversion asymmetry)[19] and the Dresselhaus effect (bulk inversion asymmetry)[20]. As a result, the spin vector of each electron precesses around an effective magnetic field. This precession is randomized by inter-subband scattering between different subbands that have different Dresselhaus interaction strengths[14]. As a result, the *ensemble averaged spin* decays with time resulting in D'yakonov-Perel' type relaxation.

We have considered a case where $E_x$ = 2kV/cm and the lattice temperature T =30K. The details of the simulation approach (which is based on a Monte Carlo simulator modified to study spin transport) can be found in reference 15 and will not be repeated here. In the simulation, we consider only the D'yakonov-Perel' relaxation[17] and ignore the Elliott-Yafet[18], Bir-Aronov-Pikus[21] and all other relaxation mechanisms (including relaxation due to hyperfine interactions with the nuclei[22]), since these are insignificant compared to the D'yakonov-Perel' relaxation in the present case.

In Figure 2, we show that the ensemble average spin component along the wire axis $\langle S_x \rangle(t)$ decays to zero after 6 ns and thereafter continues to fluctuate around zero, signaling the onset of complete depolarization. We will study the nature of this spin fluctuation.

We define spin autocorrelation function as follows:

$$C(\tau) = \int_{t_0}^{\infty} \left[ \langle S_x \rangle(t) - \langle S_x \rangle_{av} \right] \left[ \langle S_x \rangle(t+\tau) - \langle S_x \rangle_{av} \right] dt \qquad (1)$$

where

$$\langle S_x \rangle_{av} = \int_{t_0}^{\infty} \langle S_x \rangle(t) dt \tag{2}$$

and the variable τ is generally referred to as "delay time".

We observe from Figure 2 that $\langle S_x \rangle(t)$ varies randomly around zero for $t \geq t_0$ where $t_0$ is the time taken to reach steady state. Hence, $\langle S_x \rangle_{av} = 0$.

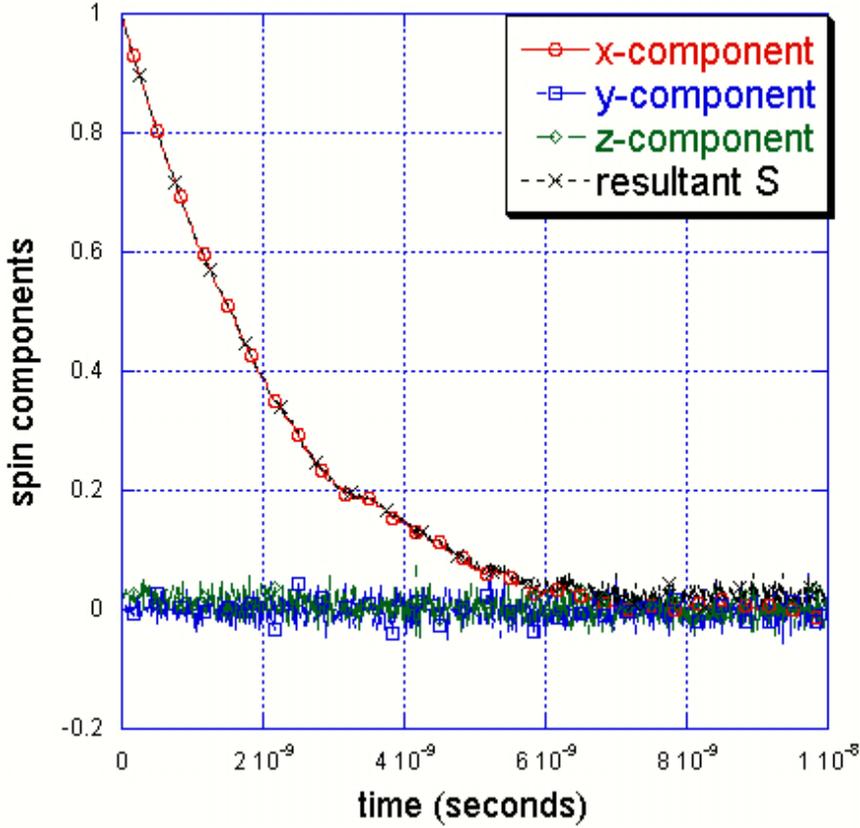

Figure 2. Temporal dephasing of the x, y and z components of ensemble average spin in the GaAs quantum wire at 30K. The driving electric field is 2kV/cm and the spins are injected with their polarization initially aligned along the wire axis (x-axis).

The noise spectral density is defined as the cosine transform of the autocorrelation function and is expressed as follows:

$$S(f) = \int_{0}^{\infty} [C(\tau) \cos(2\pi f \tau)] d\tau \tag{3}$$

3. **Results and discussion**

Figure 3 shows the autocorrelation function of the spin fluctuations at a driving electric field of 2kV/cm and the lattice temperature of 30K. The autocorrelation function decays rapidly and becomes almost zero for τ = 0.375 ns. Beyond this point it shows very small fluctuation around zero.

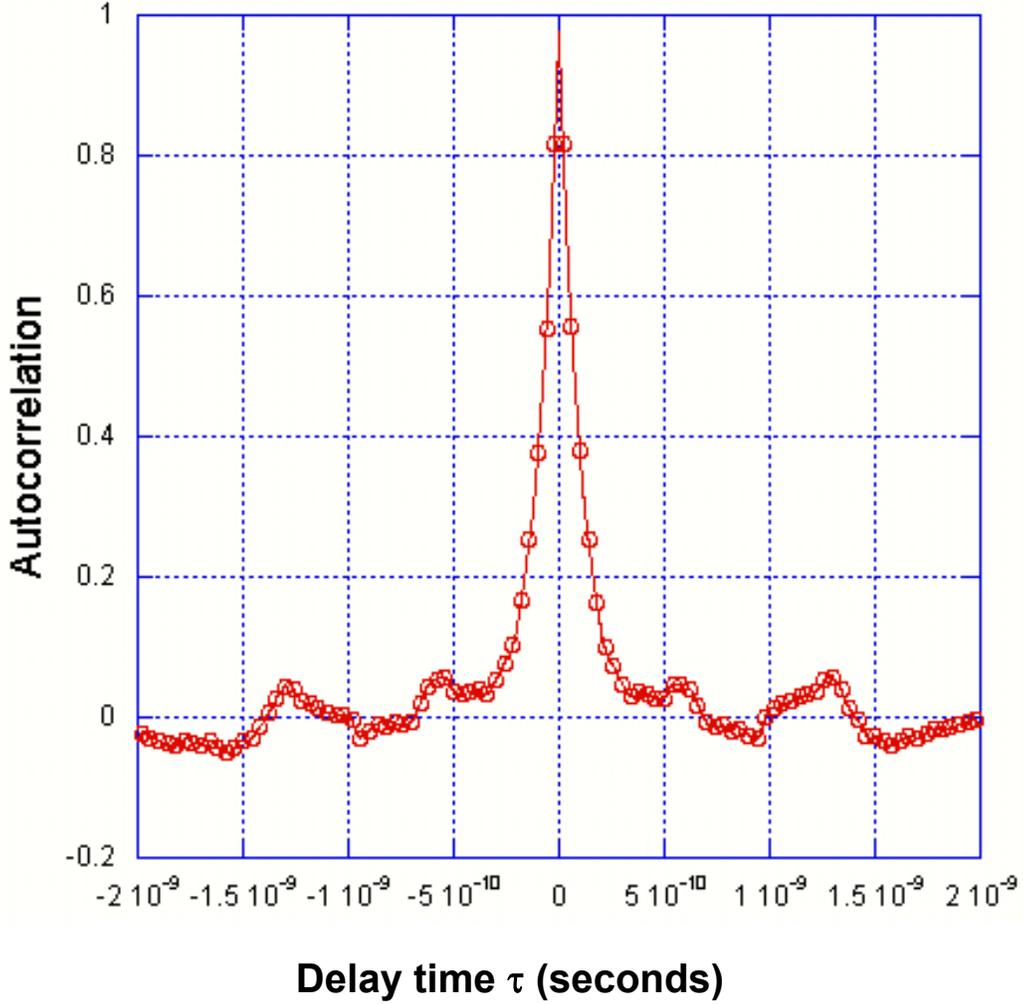

Figure 3. Autocorrelation function of the spin fluctuations in the GaAs quantum wire at a driving electric field of 2 kV/cm and at a lattice temperature of 30 K.

The associated noise spectral density is shown in Figure 4. It decays rapidly within 10 GHz.

**4. Conclusion**

In this paper we have studied, for the first time, "spin noise" in a semiconductor structure using a semi-classical approach. The autocorrelation function has no long-duration component indicating that once steady state is reached, there is no long-lived "memory" of the initial spin state in the fluctuations. The D'yakonov-Perel' relaxation is therefore an efficient relaxation mechanism that completely erases any long-lived memory of the initial spin state.

**5. Acknowldegement**

This work is supported by the US National Science Foundation under grant ECS-0196554. The authors are indebted to Prof. Marc Cahay for stimulating discussions.

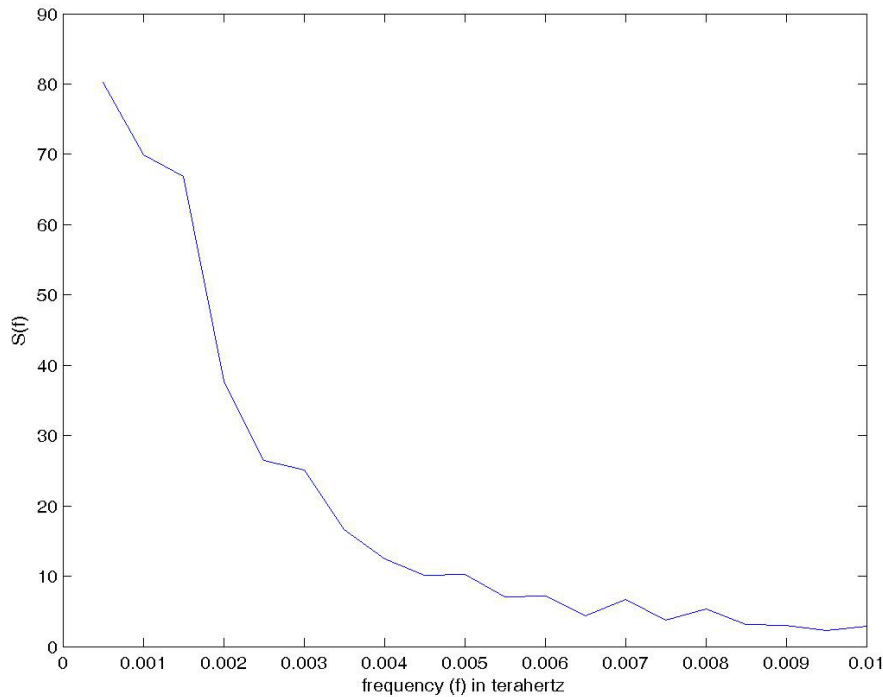

Figure 4. The spectral density of "spin-noise" in the quantum wire for driving electric field=2kV/cm and lattice temperature=30K.